# Multi-Connectivity In Mobile Networks: Challenges And Benefits

Carlos Pupiales, Daniela Laselva, Quentin De Coninck, Akshay Jain, and Ilker Demirkol

*Abstract*—**Satisfying the stringent 5G Quality of Service (QoS) requirements necessitates efficient resource utilization by the mobile networks. Consequently, we argue that Multi-Connectivity (MC) is an effective solution to leverage the limited radio resources from multiple base stations (BSs) in order to enhance the user throughput, provide seamless connectivity, or increase the data reliability. For this, we study different MC architectures, where distinct network entities and protocol layers are used to split or aggregate the user traffic. The benefits and challenges of MC are analyzed as well as the open issues that network/device vendors and mobile network operators (MNOs) have to address for its use. Finally, through experimental evaluations, we illustrate the importance of MC design decisions for the overall network performance.**

## I. Introduction

Every new generation of mobile networks is expected to support a broader range of services with distinct key performance indicators (KPIs). For instance, 5G critical services require very low latency and a high level of reliability, whereas the broadband services demand support for high traffic density per unit area and high data rates per user [1]. To support these diverse QoS requirements, it is critical to use the spectrum resources efficiently. For instance, several approaches to improve the user data rate can be considered, such as increasing the bandwidth, improving the spectral efficiency, implementing cell densification, and enabling effective coordination between multiple BSs [2]. Nevertheless, the spectrum resources are scarce and costly, the hardware complexity and its cost limit the spectral efficiency, and the cell densification results in higher interference along with higher capital and operating expenditures.

Carrier Aggregation (CA), where the User Equipment (UE) consume radio resources of the same BS and same Radio Access Technology (RAT), has been widely used to improve the user throughput. However, it is still limited by the scarcity of the bandwidth resources assigned to a BS. For this reason, Multi-Connectivity (MC) emerges as an alternative solution that allows the MNO to leverage bandwidth resources from different BSs to enhance the user performance. Indeed, these technologies complement each other, and they can harmoniously coexist. In this article, we define MC as a broad concept, where the UE can simultaneously consume radio resources of multiple BSs operating same or different RATs such as Long Term Evolution (LTE), 5G New Radio (NR), and Wi-Fi.

MC can play an essential role in achieving several KPI targets defined for the typical 5G use case scenarios: Ultra-Reliable Low Latency Communications (URLLC), enhanced Mobile Broadband (eMBB), and massive Machine Type Communications (mMTC) [1]. For instance, MC can flexibly attain improved data rate through traffic aggregation for eMBB or improved data reliability through path redundancy for URLLC. Such objectives can be achieved at reasonable implementation costs, yet different MC designs may entail different levels of coordination, and hence different deployment and operation complexity.

This paper analyses different MC architectures options, which could improve the user throughput, increase the data reliability, and provide a seamless connection. We detail the advantages provided by MC, along with the open research issues that are critical to achieving an efficient MC operation. In addition, we review the standardized MC solutions and compare the benefits and limitations they provide.

Furthermore, to show the challenges that MC operation brings to the overall system performance, we implement and evaluate an MC solution using software-defined radio based implementations for the eNB and UE. We showcase the impact of MC design decisions and their importance on end-to-end system performance. For this, we study the throughput performance of saturated TCP and UDP traffic using the single-connectivity (SC) and MC approaches on a mobile network testbed, in the meantime quantifying the effect of a critical MC design parameter.

## II. Multi-connectivity Architectures

In any MC design, the user plane (UP) traffic can be split or duplicated at a given transmitting protocol layer and then reversely aggregated or assembled in the mirrored receiving protocol layer. The protocol layer mentioned above, denoted as *MC anchor layer* in this article, and the layers above are shared by protocol stacks formed by lower layers, denoted as *communication paths* in this work. Different MC architecture options can be envisioned to enable MC operation, wherein the MC anchor layer can be located at the core network (CN), radio access network (RAN), or application server (end-to-end). In the following, we present such architectures, along with several MC standardized solutions.

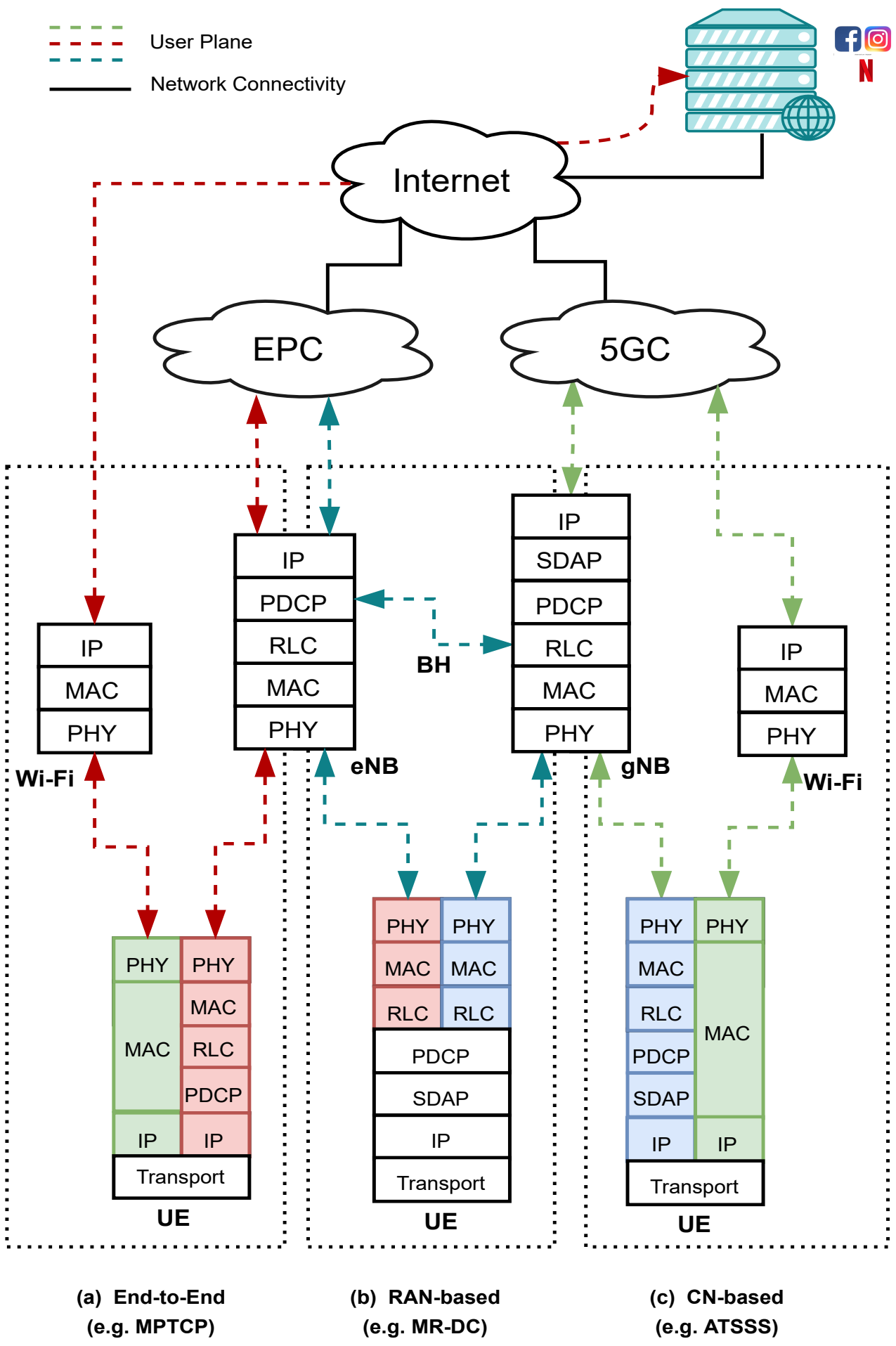

Fig. 1. MC architecture examples from the user plane perspective.

## A. RAN-based MC

In RAN-based MC, radio resources from multiple radio paths are coordinated among multiple BSs for a given UE. Depending on the MC solution, BSs might communicate using a specific interface through the backhaul (BH) (e.g., 5G BSs communicate through the Xn interface). In contrast, if at least one LTE BS is in use, the X2-U interface is used. In this case, one of the BSs acts as an anchor entity to host the MC anchor layer for the UP and to manage the MC-related CP procedures regardless of the CN configuration. Fig. 1(b) illustrates the UP implementation for a RAN-based MC solution.

The anchor BS uses the Radio Resource Control protocol to exchange signaling messages among the BSs to maintain the MC operation with the UE. Note that radio-specific functionalities for each communication path are handled independently by each BS. Additionally, having the MC anchor layer in the RAN helps the radio resource management (RRM) algorithms quickly respond to sudden channel quality changes. However, the hardware and/or software upgrade costs required for every BS may limit the scalability of MC deployments.

Although the communication between the anchor layer and the communication paths located on different BSs seems to be similar to that of functional splits [3] (e.g., F1 interface defined by 3GPP) the messaging and the context information required for MC are different. Yet, similar to the functional split options, the communication between the entities mentioned above might define latency and/or capacity requirements for the backhaul that may challenge the effectiveness of MC solutions.

In the following, possible MC anchor layers for the RAN-based MC architecture option are explained.

*1) Physical (PHY) Layer:* In PHY-MC, the use of spectrum and/or time resources from multiple BSs are coordinated for a given UE. The UE's interference information is used to decide the BSs to communicate with the UE at a given time. For PHY-MC to be efficient, a continuous control information exchange is necessary between the UE and BSs, and among BSs. Such a level of coordination defines stringent latency requirements, especially for backhaul links. Moreover, complex solutions are required to support different numerologies, frame structures, and waveforms potentially used by different BSs.

Multi-Transmission and Reception Points (multi-TRP), in 5G, and Coordinated Multi-point (CoMP), in 4G, are MC technologies, which coordinate the transmission and reception of user data from several BSs in order to mitigate the inter-cell interference. However, the aforementioned challenges limit their effective use [4], [5].

*2) Medium Access Control (MAC) Layer:* In MAC-MC, the transmission of transport blocks (TBs) is coordinated with multiple PHYs. The main benefit of MAC-MC is the fast traffic switching between communication paths thanks to the rapid system adaptation in case of radio link changes.

Nevertheless, supporting in a single MAC entity different transport block sizes, time slot durations, and Hybrid Automatic Repeat Request (HARQ) procedures, defined for the RATs, will considerably increase the scheduler complexity. For instance, NR uses a codeblock-based asynchronous HARQ for uplink and downlink, while LTE uses TB-based synchronous HARQ for UL and asynchronous for downlink. Furthermore, when the PHY and MAC reside on different locations (e.g., different BSs) achieving the smaller slot durations envisioned for URLLC applications becomes challenging.

5G NR in Unlicensed Spectrum (NR-U) with CA can be considered a MAC-MC solution since it uses a license-assisted access (LAA) method to transmit user traffic via different RATs located at the same BS, using licensed and unlicensed spectrum, respectively. [6]. In this case, the MAC from the NR stack controls or coordinates both communication paths.

*3) Packet Data Control Protocol (PDCP) Layer:* PDCP is suitable to act as an MC anchor layer since its procedures do not face very tight timing constraints as MAC or PHY do and because its specifications are similar for LTE and NR. Hence, each communication path can perform its link adaptation and resource allocation procedures. In this sense, it is possible to increase the user throughput or the data reliability by splitting/aggregating or duplicating a packet data unit (PDU) using multiple BSs without incurring major implementation complexities [7]. The following 3GPP technologies can be considered as PDCP-MC solutions:

*a) Dual Connectivity (DC) and Multi-Radio Dual Connectivity (MR-DC):* The UE is simultaneously served by two 3GPP BSs, with the same RAT for the former and heterogeneous RATs for the latter solution. In both cases, the MC-related CP aspects are handled by one of the two BSs, while the UP uses one or both BSs simultaneously [2], [8].

*b) NR-U with DC:* UE can communicate simultaneously with two 3GPP BSs, the RATs of which operate on licensed and unlicensed spectrum, respectively [6]. An anchor LTE or NR BS manages the NR-U BS and the MC user and control planes.

*c) NR Dual Active Protocol Stack (DAPS):* The UE is simultaneously connected to two NR BSs to ensure a seamless handover. Unlike 3GPP-defined DC solutions, in DAPS, there is no anchor BS.

*d) LTE-WLAN Aggregation (LWA):* It integrates LTE and Wi-Fi networks controlled by the same MNO. The eNB decides whether offload the traffic to Wi-Fi or aggregate it between the LTE and Wi-Fi networks.

*4) Network Layer:* It can act as the MC anchor layer since the IP protocol is the de facto network layer protocol for all 3GPP and non-3GPP networks. The MC operation between 3GPP and non-3GPP BSs requires an IP tunnel between the UE and the 3GPP BS for security and encapsulation purposes. Moreover, different IP addresses should be configured at the UE to route the IP packets. However, MNOs usually configure a single IP address per UE in their CN. Therefore, new functionalities in the IP header may be required to support traffic aggregation (e.g., packet sequence identification). These functionalities bring implementation complexities and additional costs that may limit the MC deployments.

LTE WLAN Radio Level Integration with IPsec Tunnel (LWIP) is a Network-MC solution, where data transfer from LTE to WLAN and vice versa is done using an IPSec tunnel. LWIP does not require any changes on the WLAN infrastructure, unlike the LWA. The eNB manages the MC-related CP and UP functionalities, while each BS handles its radio-specific functionalities.

### *B. CN-based MC*

In this MC architecture option, the CN directly manages the MC-related CP and UP functionalities. For this, BSs that are connected to the same or different CNs can serve the UE. The latter case requires coordination between both entities (e.g., between LTE and 5G CNs) for which new communication procedures are needed. However, in such a scenario, a RAN-based MC approach results in a simpler alternative. In both cases, disjointed communications paths are required to transport the user traffic through multiple BSs. This can be done using independent CN connections for each BS, and thus, different IP addresses at the UE. Even though this approach is an affordable solution that offers versatility and scalability regardless of the number of BSs and their technology, it also imposes challenges for MNOs. For instance (i) implementation of new capabilities to split and aggregate UP traffic that belongs to different packet data networks; (ii) quickly adapt to the dynamic radio link conditions since the CN typically has no information about that.

In this regard, 3GPP introduced the *Access Traffic Steering, Switching, and Splitting (ATSSS)* technology that gives the MNO the control to steer UP traffic between NR and Wi-Fi networks. ATSSS uses specific multipath functions between the CN and UE to transport the user traffic and CP messages, such as traffic and round trip time measurements. Currently, ATSSS only supports TCP-based traffic with the use of the Multipath TCP (MPTCP) protocol. Fig. 1(c) illustrates the UP implementation for ATSSS.

### *C. End-to-End MC*

In this MC architecture option, MC operation is enabled using capabilities of upper protocols between multiple 3GPP BSs or between 3GPP and non-3GPP BSs. In the former case, the UE and the CN can establish a redundant UP connection in an end-to-end fashion for reliability purposes [9]. This solution is relevant for URLLC services. However, it comes at higher deployment costs since it needs redundancy in all network entities (i.e., RAN and CN). In the latter case, MC operation is established between the UE and the application server. Hence it is entirely agnostic to the RAT and CN configuration. In fact, this is the simplest approach to enable MC operation. However, it is not possible to guarantee a given QoS criteria since the MC operation is not in the control of the MNO. In both cases, *Transport layer* protocols can be exploited to provide such functionalities. Fig. 1(a) shows the MC operation between LTE and Wi-Fi networks.

In this regard, MPTCP and Multipath QUIC (MPQUIC) protocols are affordable options to enable MC operation and transport TCP-based traffic. In this sense, independent traffic flows, one per communication path, can be created using different IP addresses and/or port numbers [10], [11].

## III. Benefits and Limitations

In this section, we describe the benefits that MC offers and the limitations encountered to provide an effective MC operation. These aspects are summarized in Table I for the standardized MC solutions.

### *A. Improved Data Rate*

In MC operation, the UE can combine multiple data streams from different BSs into a single data stream to enhance the user data rate. Ideally, the resulting aggregate throughput would be equal to the sum of the throughputs of the communication paths used. Nevertheless, the different radio link conditions and assigned radio resources experienced at each BS and the delay difference between both communication paths can cause out-of-order packet arrivals, negatively affecting the performance of the upper layer protocols, such as TCP. Indeed, the obtained MC throughput can even be lower than the one achieved in SC as showed in [12]. Additionally, the chosen MC solution might achieve application-level improvements depending on the network scenario. For instance, MPTCP and MPQUIC can increase the throughput if they can use disjointed communication paths.

TABLE I
BENEFITS AND LIMITATIONS OF MC STANDARDIZED SOLUTIONS

| Anchor Layer | Standardized Solution | Technical Specification | Technical Objective | Main Limitation |
|---|---|---|---|---|
| PHY | CoMP | TS 36.300 | Improvements in SINR | Requires high level of coordination between BSs |
| | Multi-TRP | TS 38.300 | | |
| MAC | NR-U CA | TS 38.331 | Rapid system adaptation in case of link failures | Complex packet scheduler |
| PDCP | DC | TS 36.300 | Higher data reliability; Higher throughput | Requires additional hardware and software capabilities at the UE |
| | MR-DC | TS 37.340 | | |
| | LWA | TS 36.300 | Higher throughput | Aggregation limited to LTE and Wi-Fi controlled by the MNO |
| | NR-U DC | TS 38.331 | Higher throughput | Requires additional hardware and software capabilities at the UE |
| | DAPS | TS 38.300 | Mobility robustness | Currently available between 5G BSs |
| Network | LWIP | TS 36.300 | Deployment affordability | Traffic offload limited to LTE and Wi-Fi |
| Transport (End-to-End) | MPTCP | RFC 6824 | Higher throughput; Deployment affordability | MC operations are not in the control of the MNO |
| Transport (Core Network) | ATSSS | TS 24.193 | Higher throughput; Higher data reliability | MC operation limited to TCP-based traffic |

### B. Improved Reliability

In mobile networks, reliability is typically provided by retransmitting the erroneous data. Despite being effective, it is also time consuming. Hence, this procedure is not suitable to simultaneously meet the reliability and low latency requirements needed by URLLC applications. MC solutions can be exploited to provide reliability and low latency by sending redundant data using different BSs. In this regard, the Packet Duplication (PD) feature, defined for MR-DC, can satisfy the high reliability and low latency requirements, where identical control or data packets are sent through multiple BSs [7].

### C. Mobility Robustness

MC solutions can reduce the interruption time, and the amount of signaling required for the seamless connection envisioned for future networks. The UP traffic can be rapidly switched from one BS to another through the backhaul link, thus avoiding the intervention of the CN. For instance, the master BS provides the CP functionality to the UE in MR-DC. Hence, the UP is switched from the master to the secondary BS during a handover without involving the CN. This process is faster and offers lower signaling overhead compared to the traditional handover [2]. Even though one radio interface is sufficient for UP traffic, RRM procedures by both BSs are needed, so the UE must use both radio interfaces actively, like in the DAPS case. Similarly, MPTCP and MPQUIC can provide mobility robustness. However, in that case, the UE mobility is only possible between 3GPP and non-3GPP networks [10], [11].

### D. Service Segregation

MC can be used to serve a UE requesting services with distinct requirements by segregating those services to different communication paths. For instance, the UE can be connected to two BSs whose RATs use mmWave and Sub-6 GHz carriers, respectively. The former can be used for services that demand high throughput (e.g., video streaming) while the latter can be used for lower throughput, such as mMTC. This feature may be suitable for emerging applications, which define flows with different QoS criteria such as V2X.

### E. Deployment Cost Savings

In order to address the massive increase in throughput and the number of UE connections envisioned for eMBB applications, the MNOs have to extensively deploy 5G capabilities, especially in heavily populated areas. This deployment cost can be alleviated adopting RAN-based MC solutions, as the one used for the 5G non-standalone mode, instead of massively deploy new costly RAN and CN infrastructures. This can be a cost-effective and scalable approach to reduce the time to market and bring the 5G capabilities to scenarios where to completely deploy 5G infrastructure is not profitable (e.g. in rural areas). This strategy can also be used to satisfy temporary demands (e.g. for concerts or stadiums).

### F. MC Messaging Overhead

For PHY-MC and MAC-MC solutions, the radio resource allocation, done at each time slot duration, brings stringent requirements that increase the signaling traffic and limit the allowed backhaul delay. For instance, PHY-MC requires that the data for the UE is processed and forwarded to the corresponding BSs within one slot duration [4]. For MAC-MC, the scheduling control and HARQ synchronization required by the MAC scheduler to serve multiple PHYs, limit the backhaul delay to 2 ms [3].

Contrarily, PDCP-, Network-, and CN-based MC require less frequent data usage reports, the frequency of which depends on the flow control method in use. Although these reports do not impose specific delay constraints for the backhaul, a considerable delay may affect the performance of transport and application layers (e.g., because of TCP's retransmissions by timeout). In general, how often MC control messages are exchanged should be decided based on the variability of the radio link conditions. However, the overhead created by these messages increases linearly with the frequency they are exchanged, along with the number of UEs and BSs involved in the MC operation.

## IV. Challenges and Open Issues

Although MC offers several benefits as indicated in Section III, there are several challenges and open issues that still need to be addressed. Table II summarizes the relevant open research questions that need further investigation.

### *A. Packet Reordering*

Different radio link conditions, RAT procedures, and communication path delays can cause packets to arrive out of order. Hence, a packet reordering solution is needed to avoid a negative impact on the performance of the upper layer protocols (e.g., for TCP). The 3GPP reordering method defined for DC and MR-DC [13] uses a static reordering timeout, the value of which must be carefully chosen. If not, packets can be excessively buffered, creating issues, such as bufferbloat, thus degrading the performance of time-sensitive applications. A suitable reordering timeout value should be chosen considering aspects such as backhaul latency, radio link conditions, traffic type, buffer length, and QoS requirements.

### *B. Dynamic Flow Control*

An inadequate flow control logic might create under- or over-utilized links, resulting in out-of-order packet arrivals and poor overall system performance. For instance, a static flow control logic is inefficient since the data rate from each BS is spatio-temporal (i.e., it depends on the instantaneous radio link conditions and assignment of radio resources). If such aspects are not considered, it may be necessary to buffer the PDUs at the sender and/or reorder them at the receiver. Several factors rule the flow control logic such as the backhaul latency, radio link conditions, and QoS requirements.

### *C. Cross-layer Design*

Flow control and reordering algorithms can make better decisions by considering the information from the different protocol layers, such as the application requirements, the transport protocol used, and the network conditions. For instance, unlike UDP, TCP has to ensure an in-sequence delivery, which might affect the application performance due to the delays and retransmissions it might incur. A careful design of the packet reordering at the receiving MC anchor layer can reduce such problems and improve the application performance. However, an inefficient MC reordering mechanism can result in additional delays that could cause spurious TCP retransmissions.

### *D. MC Operation Management*

The MNO should decide when to use MC instead of SC and which BSs should be involved in this MC operation. For instance, inefficient MC decision, user association, and resource allocation methods can degrade the performance of some UEs and even the overall system performance. For the decision-making, the MNO can consider the user QoS requirements, the terminal capabilities, and the spatio-temporal network KPIs. Such decisions can be improved by collecting and processing MC operation data through reinforcement learning and data analytics techniques.

### *E. Beyond Two RATs*

Since recent UEs are already equipped with 4G, 5G, and Wi-Fi interfaces, they can use more than two BSs simultaneously to transfer the UP traffic. One of the benefits of this MC approach is the versatility of aggregating traffic even though one BS fails. This approach is not possible with the current MC standardized solutions since they consider only two BSs. Nevertheless, this new approach may increase the flow control and reordering algorithms' implementation and management complexity.

## V. MC Challenge Showcase

To quantitatively illustrate the importance of efficient *Packet Reordering* and *Dynamic Flow Control* methods for the performance of the upper layer protocols, we implement and evaluate DC on an LTE testbed. This PDCP-MC solution has been chosen since it is the preferred method for the first commercial deployments of 5G through the EN-DC solution [8]. For our evaluations, DC with the split bearer option is implemented using the LTE/5G-NR compliant Open Air Interface software for eNB and UE. Further details of the implementation and testbed configuration can be found at [12]. We study the performance of TCP and UDP protocols for downlink traffic in three scenarios: (i) single connectivity (SC), (ii) DC without packet reordering function (DC_NoR), and (iii) DC with packet reordering function (DC_Reo) using the 3GPP algorithm given in [13].

Since 3GPP does not specify a concrete value for the reordering timeout, values of 40, 60, 80, 100, and 150 ms have been evaluated. Moreover, a simple flow control logic based on a round-robin traffic distribution is used at the master eNB. For this, the master eNB, denoted as SC_A, and the secondary eNB, denoted as SC_B, are connected using the X2-U interface, which adds a fixed delay of 10 ms. Furthermore, for realistic analysis and to demonstrate that the aggregate throughput is affected by the variance in channel capacities, we use a radio link channel trace from a pedestrian user [14]. In this sense, each eNB uses different channel quality indicator (CQI) sets, the values of which change every second and produce different throughput results. The throughput obtained in the SC scenario, illustrated in Figs. 2 and 3, serve as a baseline to benchmark the aggregate throughput obtained in the DC cases. For both TCP and UDP analysis, the traffic is generated using the *iperf3* tool in sessions of 30 seconds.

TABLE II
SUMMARY OF OPEN RESEARCH QUESTIONS

| | |
|---|---|
| Packet Reordering | • How to handle packet reordering for a subset of the traffic when multiple traffic flows exist?<br>• How long should a packet wait in the reordering buffer without affecting the communication performance? |
| Flow Control | • How to dynamically choose the splitting ratio considering the performance of the upper layer protocols? |
| Cross-layer Design | • How MC-based algorithms exploit the upper layer protocol information? |
| MC Operation Management | • Which is the entity in charge of managing the MC-related procedures?<br>• What are the KPI targets to be considered for MC decisions? |
| Beyond Two Rats | • Is it possible to effectively aggregate data streams from more than two sources? |

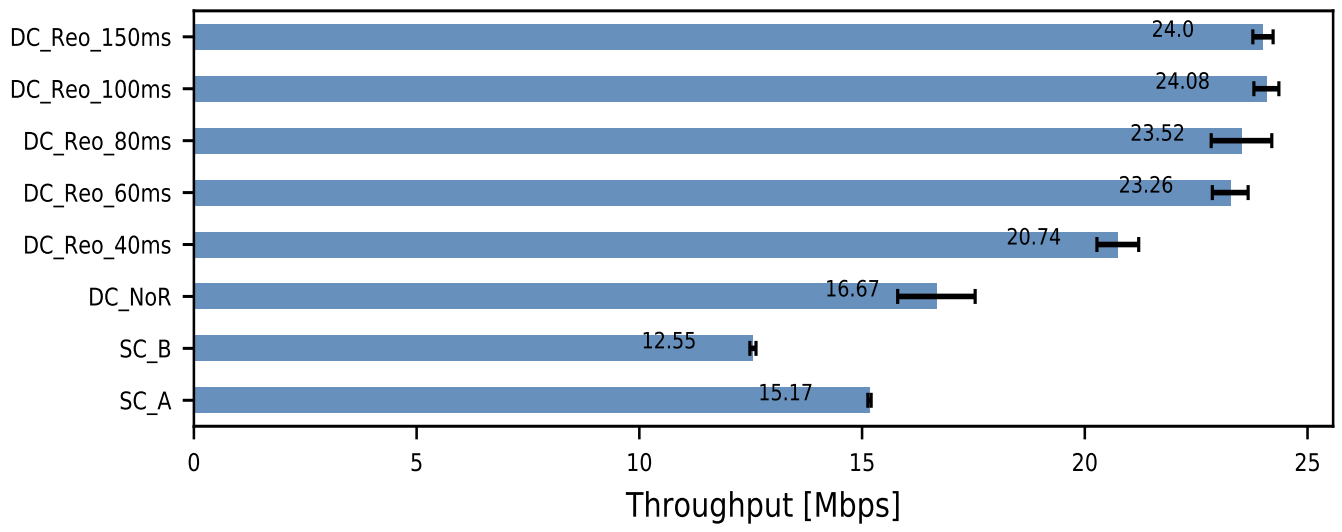

Fig. 2. Throughput analysis for TCP traffic.

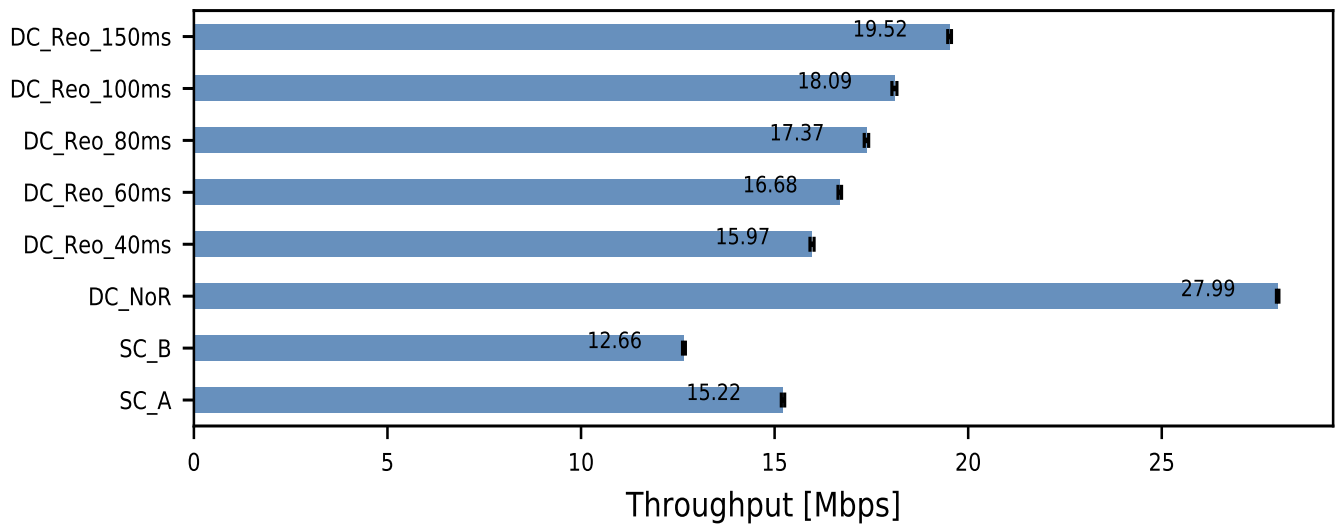

Fig. 3. Throughput analysis for UDP traffic.

For TCP traffic, the results presented in Fig. 2 for the DC_Reo case show that the aggregate throughput is on average 36% to 58% higher than the one achieved by SC_A. However, it is lower than the ideal aggregate throughput (i.e., 27 Mb/s). Similarly, in the DC_NoR case, the aggregate throughput is only 9.8% higher than the one achieved by SC_A. In the latter, the out-of-order packets make TCP go continuously into the congestion avoidance phase. Hence, causing fast retransmissions and/or retransmissions that degrade the throughput performance.

Additionally, in the DC_Reo case, the aggregate throughput depends on the chosen reordering timeout value. If this value is not enough to compensate for the delay differences between the communication paths, out-of-order packets are delivered to the upper layers. On the contrary, a large reordering timeout may cause the packets to wait excessively in the PDCP reordering buffer. This issue increases the end-to-end delay, which can trigger TCP retransmissions by timeout. In both cases, an inadequate timeout value negatively affects the throughput performance.

For UDP traffic, as illustrated in Fig. 3, the ideal aggregate throughput is achieved in the DC_NoR scenario (i.e., 27 Mb/s). However, when the reordering function is in use (i.e., DC_Reo case) the obtained throughput is lower than the one obtained in DC_NoR. Since UDP is not a reliability-oriented protocol, it always sends a fixed amount of data controlled by the application. This behavior, along with the flow control logic used in this experiment, results in most packets arriving at the UE through the faster communication path. Hence, placing them in the PDCP reordering buffer is necessary until the delayed ones arrive or the reordering timeout expires. Packet reordering can reduce the throughput. Nevertheless, for some UDP-based applications, where the out-of-order packets degrades the application performance (e.g., in a video conference) a delay-limited packet reordering can be beneficial.

For both TCP and UDP, the results show that the MC aggregate throughput is significantly influenced by how the differences in terms of latency, radio link conditions, and channel capacity between communications paths are managed by the flow control and packet reordering mechanisms. Indeed, all types of MC solutions' performances are expected to be affected in varying degrees due to the differences mentioned above. DC can significantly boost the throughput for UDP traffic but at the cost of delivering out-of-order packets, which can affect the performance of applications. Similarly, TCP performance can be improved by the packet reordering process but with a benefit that depends on the reordering timeout value.

## VI. CONCLUSIONS

In this article, we have discussed the opportunities and challenges that MC offers to effectively utilize the system resources to enhance the user throughput, increase the data reliability, and reduce the negative effects of handover. We presented and discussed different MC architecture options, categorized by the protocol layer where the UP traffic is split or aggregated. Albeit the system enhancements MC promises, it also defines several challenges for its efficient implementation. When adopting TCP and UDP protocols, our experimental evaluations of MC and SC solutions showed that the design decisions such as the flow control and packet reordering schemes can significantly impact the overall system performance and thus should be carefully considered.

## ACKNOWLEDGMENT

This work is partially supported by the Secretaría de Educación Superior, Ciencia, Tecnología e Innovación (SENESCYT), Ecuador.

## Biographies

**Carlos Pupiales** (carlos.pupiales@upc.edu) received the B.Sc. degree from Escuela Politécnica Nacional, Ecuador, and the M.Sc. degree in telecommunications engineering from the University of Melbourne, Australia. He is currently pursuing the Ph.D. degree with the Department of Network Engineering, Universitat Politècnica de Catalunya, Spain. His main research interest includes the area of network protocols in mobile networks. He was a recipient of the Best Demo Award in IEEE MASS 2019.

**Daniela Laselva** (daniela.laselva@nokia-bell-labs.com) received her M.Sc.E.E. in 2002 from Politecnico di Bari, Italy. She is currently with Nokia-Bell Labs, Aalborg, Denmark, where she works as a senior researcher. At present, she is engaged with 5G/NR design and standardization, including multi-connectivity solutions, support of mission-critical applications, and E2E performance optimization. She is also the author/co-author of dozens of peer-reviewed publications and patents on a wide range of topics.

**Quentin De Coninck** (quentin.deconinck@uclouvain.be) received his Ph.D. from UCLouvain, Belgium in 2020 under the supervision of Prof. Olivier Bonaventure. He is now a post-doctoral researcher in the same institution. His research interests include Internet protocol design and low-level system architecture.

**Akshay Jain** (akshay.jain@neutroon.com) received his PhD from Universitat Politècnica de Catalunya, Spain in 2020. He is now the VP of Telecom Engineering at Neutroon Technologies S.L., Spain. His research interests include 5G and beyond networks, machine learning, and optimization methods.

**Ilker Demirkol** (ilker.demirkol@upc.edu), IEEE Senior Member, is an Associate Professor in the Department of Mining, Industrial and ICT Systems Engineering at the Universitat Politècnica de Catalunya, Spain, where his research currently has the focus on network algorithmics, Internet of Things and mobile networks. He received the BSc., MSc., and PhD. degrees in computer engineering from the Bogazici University, Istanbul, Turkey. Over the years, he has worked in a number of research laboratories and industrial corporations in Europe and the USA.